\documentclass[12pt]{article}
\usepackage{graphicx}

\begin{document}

\vbox{\vspace{6ex}}
\begin{center}
{\LARGE \bf  Historical Approach to Physics according to Kant, Einstein, and Hegel} \\

\vspace{5mm}

Y. S. Kim \\
Center for Fundamental Physics, University of Maryland, \\
College Park, Maryland, 20742, U.S.A.

\vspace{5ex}

\end{center}

\begin{abstract}

It is known that Einstein's conceptual base for his theory of
relativity was the philosophy formulated by Immanuel Kant.
Things appear differently to observers in different frames.
However, Kant's Ding-an-Sich leads to the existence of the
absolute reference frame which is not acceptable in Einstein's
theory.  It is possible to avoid this conflict using the
ancient Chinese philosophy of Taoism where two different
views can co-exist in harmony.  This is not enough to explain
Einstein's discovery of the mass-energy relation.  The energy-momentum
relations for slow and ultra-fast particles take different forms.
Einstein was able to synthesize these two formulas to create his
energy-mass relation.  Indeed, this is what Hegelianism is about in
physics.  Isaac Newton synthesized open orbits for comets and closed
orbits for planets to create his second law of motion.  Maxwell
combined electricity and magnetism to create his four equations to
the present-day wireless world.  In order to synthesize wave and
particle views of matter, Heisenberg formulated his uncertainty
principle.  Relativity and quantum mechanics are the two greatest
theories formulated in the 20th Century. Efforts to synthesize
these two theories are discussed in detail.

\end{abstract}

\vspace{10mm}
\noindent based on an invited presented at the 32nd Congress of the Italian
Society of Historians of Physics and Astronomy (Rome, Italy, September 2012).

\newpage
\section{Introduction}
Einstein studied the philosophy of Immanuel Kant during his earlier years.
It is thus not difficult to see he was  influenced  by Kantian view of the
world when he formulated his special theory of relativity.  It is also known
that, in formulating his philosophy, Kant was heavily influenced by the
environment of Koenigsberg where he spent eighty years of entire life.
The first question is what aspect of Kant's city was influential to Kant.
We shall start with this issue in this report.

\par
In Einstein's theory, one object looks differently to moving observers with
different speeds. This aspect is quite consistent with Kant's philosophy.
According to him, one given object or event can appear differently to observers
in different environments or with different mindsets.  In order to resolve
this issue, Kant had to introduce the concept of "Ding-an-Sich"   or thing
in itself meaning an ultimate object of absolute truth.  Indeed, Kant had
a concept of relativity as Einstein did, but his Ding-an-Sich led to the
absolute frame of reference.  Here Kantianism breaks down in Einstein's
theory.  Kant's absolute frame does not exist according to Einstein.
In order to resolve this issue, let us go to the ancient Chinese philosophy
of Taoism.  Here, there are two different observers with two opposite
points of view.   However, this world works when these two views form a
harmony.  Indeed, Einsteinism is more consistent with Taoism.  The
energy-momentum relations is different for a massive -slow particle and
for a fast-massless particle.  Einstein's relativity achieved the harmony
between these two formulas.

\par
This leads us to Hegel's approach to the world.  If there are two opposite
things, it is possible to derive a new thing from them.  This is what Einsteinism
is all about.  Einstein derived his $E = mc^2$ from two different expressions of
the energy-momentum relation for massive and massless particles.

\par
Einstein thus started with Kantianism, but he  developed  a Hegelian approach
to physical problems.  Indeed, this encourages us to see how this Hegelianism
played the role in developing new laws of physics. For instance, Newton's
equation of motion combines the open orbits for comets and the closed orbits
of planets.

\par
If  this Hegelian approach is so natural to the history of physics, there is
a good reason.  Hegel derived his philosophy by studying history.  Hegel
observed that Christianity is a product of Jewish one-God religion and
Greek philosophy.  Since Hegel did not understand physics, his reasoning was
based on historical development of human relations.  It is thus interesting
proposition to interpret Hegel's philosophy using the precise science of
Physics.

\par
In Secs.~2 and~3, we review how Kantianism was developed  and how
Einstein was influenced by Kant.  In Sec.~4, it is pointed out that
Hegelianism is the natural language in understanding physics.  In Sec.~5,
we examine whether quantum mechanics and relativity can be combined into
one theory according to Hegelian approach to the history of physics.

\par
\section{Geographic Origin of Kantianism and Taoism}
Immanuel Kant  (1724-1804) was born in the East Prussian city of Koenigsberg,
and there he spent 80 years of his entire life.   It is agreed that his
philosophy was influenced by the lifestyle of Koenigsberg.

\par
The city of Koenigbeerg is located at the Baltic wedge between Poland
and Lithuania.  As shown in Fig.~1, this place served as the traffic
enter for maritime traders in the Baltic Sea.  In addition, this city is
between the eastern and western worlds (Applebaum 1994).  However, there
are no natural boundaries such a rivers or mountains.  Thus, anyone with a
stronger army could come to this area and run the place.

\par

\begin{figure}[thb]
\centerline{\includegraphics[scale=0.45]{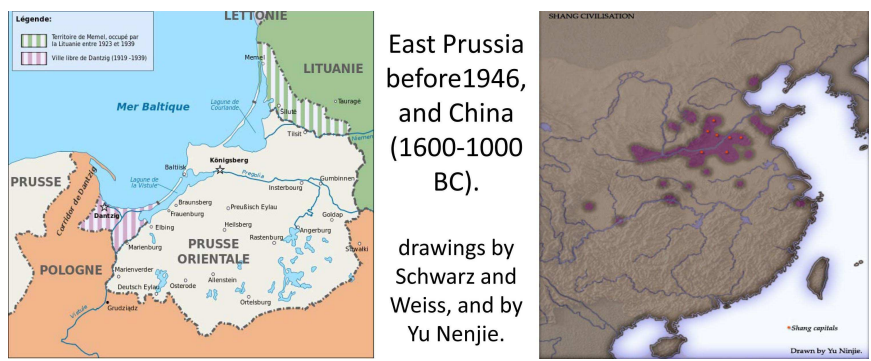}}
\caption{Koenigsberg was a German city along the Baltic coast between
 Poland and Lithuania.  It used to be a meeting place for many people
 with   different viewpoints.  This place is now a Russian city
 called Kaliningrad.  China became one country as a collection of
 many people from different population pockets with different ideas
 and different lifestyles.}\label{fig.1}
\end{figure}


\par
Indeed, Koenigsberg was a meeting place for many people with different
ideas and different view points.  Kant observed that the same thing
can appear differently depending on the observer's location or state
of mind.
\par

The basic ingredients of Taoism are known to be two opposite elements
Yang (plus) and Ying (minus).  This world works best if these two
elements form a harmony.  However, the most interesting aspect of
Taoism is that its geographic origin is the same as that of Kantianism.
Let us look at the map of China given in Fig 1during the period from
1600 to 100 BC.  China started as collection of isolated pockets of
population. They then came to banks of the Yellow River, and started
to communicate with those from other areas. They drew pictures for
written communication leading eventually to Chinese characters.

\par
How about different ideas?  They grouped many different opinions
into two groups, leading to the concept of Yang and Ying.  Immanuel
Kant considered many different views, but he concluded that there
must be one and only truth.
\par
Indeed, Taoism and Kantianism started with the same environment,
but Kant insisted on one truth called Ding-an-Sich, while Taoism ended
up with two opposing elements (Kim 2006).

\begin{figure}[thb]
\centerline{\includegraphics[scale=1.8]{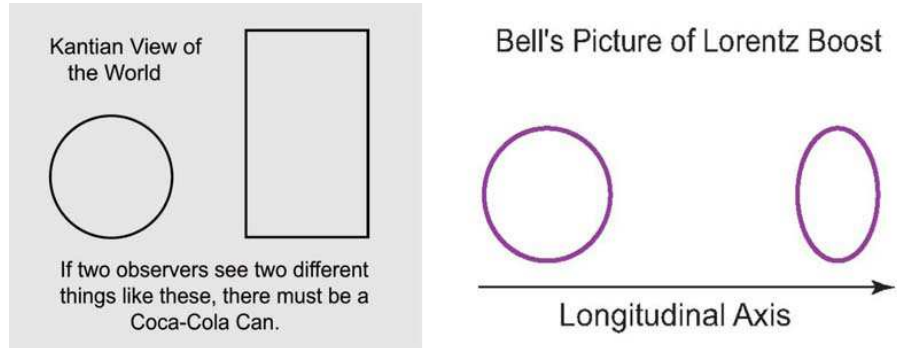}}
\caption{A Coca-Cola can appears differently to two observers
from two different angles.  Likewise, the electron orbit in the
hydrogen atom should appear differently to two observers moving
with two different speeds.}\label{fig.2}
\end{figure}


\section{Kantian Influence on Einstein}
During his early years, Einstein became quite interested in Kant and
studied his philosophy rigorously.  This was quite common among the young
students during his time.  Einstein however studied also physics, and got
an idea that one object could appear differently for observers moving
with different speeds.

\par
Let us go to Fig.~2. According Kant, an object or event look differently
to different observers depending on their places or states of mind.
A Coca-Cola can looks like a circle if viewed from the top.  It appears
like a rectangle if viewed from the side. The Coca-Cola can is an absolute
thing or his Ding-an-Sich.  Likewise, the electron orbit of the hydrogen
atom looks like a circle for an observer when both the hydrogen and the
observer are stationary. If the hydrogen atom in on a train, our first
guess is that it should look like an ellipse.   This is what Einstein
inherited from Kant.

\par

\begin{figure}[thb]
\centerline{\includegraphics[scale=0.7]{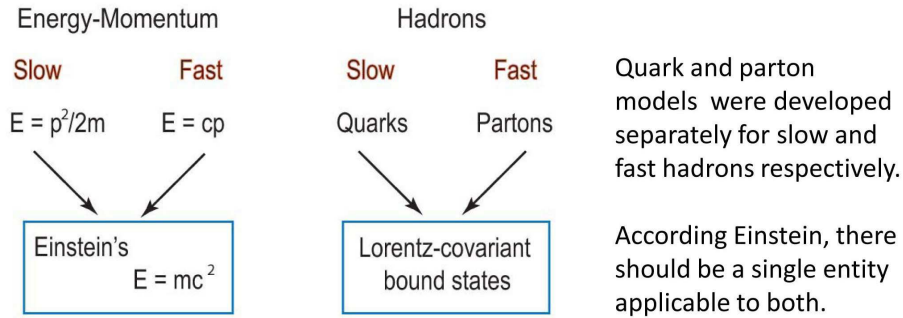}}
\vspace{-1mm}
\caption{The energy-momentum relation of a particle takes different
form when the particle moves with different speeds.   Let us choose two
limiting cases.  Einstein was able to find the same formula applicable
to both. In so doing, he found his $E = mc^2.$  Likewise, the quark model
(Gell-Mann 1964) and the parton model (Feynman 1969) should produce a
Lorentz-covariant picture of the bound state.}\label{fig.3}
\end{figure}
\vspace{2mm}

\par

However, does the hydrogen atom require a Ding-an-Sich?  The answer is No.
Indeed, Kant attempted to formulate his theory of relativity with an absolute
coordinate system corresponding to his Ding-an-Sich.  This is the basic
departure of Einsteinism from Kantianism

\begin{figure}[thb]
\centerline{\includegraphics[scale=2.0]{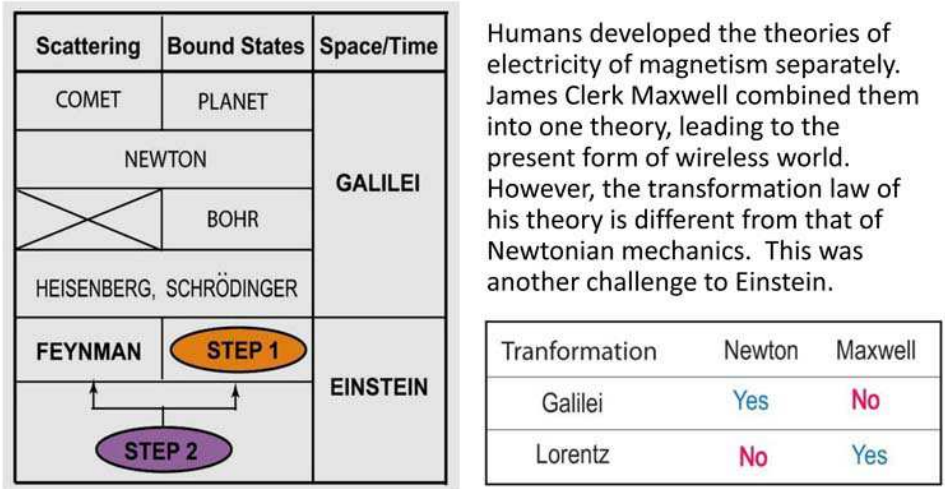}}
\caption{Progresses of physical theories are made when two theories are
combined into one.  The left side of this figure shows how mechanics was
developed.  The right figure tells how the mechanics and electromagnetism
were led to obey the same transformation law.}\label{fig.4}
\end{figure}
\par
Let us come back to Einstein.  Like Kant, Einstein started from different
observers looking at a thing differently, but ended up with a particle at
the  rest and the same particle moving with a speed close to that of light.
He then derived his celebrated energy-mass relation, as indicated in
Fig.~4 (left).  Einstein had to invent a formula applicable to both.
This is precisely a Hegelian approach to physics.  It is not clear whether
Einstein knew he was doing Hegel.  This remains as an interesting historical
problem. As for Einstein's hydrogen atom, we now have hadrons which are
bound states of quarks, while the hydrogen atom is a bound state of a proton
and  an electron.  The proton is a hadron and is a charged particle which
can be accelerated to the speed very close to that of light.  We shall
return in Sec.~5 to the problem presented in the right side of Fig.~3.
\par
However, does the hydrogen
atom require a Ding-an-Sich?  The answer is No.  Indeed, Kant attempted
to formulate his theory of relativity with an absolute coordinate system
corresponding to his Ding-an-Sich.  This is the basic departure of
Einsteinism from Kantianism Let us come back to Einstein.  Like Kant,
Einstein started from different observers looking at a thing differently,
but ended up with a particle at the  rest and the same particle moving with
a speed close to that of light.  He then derived his celebrated energy-mass
relation, as indicated in Fig.~4 (left).

\par
Einstein had to invent a formula applicable to both.  This is precisely
a Hegelian approach to physics.  It is not clear whether Einstein knew he
was doing Hegel.  This remains as an interesting historical problem.
As for Einstein's hydrogen atom, we now have hadrons which are bound
states of quarks, while the hydrogen atom is a bound state of a proton
and  an electron.  The proton is a hadron and is a charged particle which
can be accelerated to the speed very close to that of light.  We shall
return in Sec.~5 to the problem presented in the right side of Fig.~3.
\par

\section{Hegelian Approach to the History of Physics}
Since Hegel formulated his philosophy while studying history, it is
quite natural to write the history of physics according to Hegel.
First of all, Isaac Newton combined hyperbolic-like orbits for comets
and elliptic orbits for planets to derive his second-order differential
equation which  is known today as the equation of motion.

\par
James Clerk Maxwell combined the theory of electricity and that for
magnetism to formulate his electromagnetic theory leading to the
present world of wireless communication.
Max Planck observed that the radiation laws are different for
low- and high-frequency limits.  By deriving one formula for both,
he discovered Planck's constant.

\par
Werner Heisenberg observed the matter appears as a particle also
appears as a wave, with entirely different properties.
He found the common ground for both.  In so doing, he found the
uncertainty relation which constitutes the foundation of quantum m
echanics.

\par
Indeed, quantum mechanics and relativities were two most fundamental
theories formulated in the twentieth century.  They were developed
independently.  The question is whether they can be combined into
one theory.  We shall examine how Hegelian approach is appropriate
for this problem in Sec.~5.

\section{How to combine Quantum Mechanics and Relativity}
Here again, the problem gets divided into scattering and bound state problems.
Quantum field theory was developed for scattering problems, and this theory
is accepted as a valid theory, as is illustrated in Fog. 4.

\begin{figure}
\centerline{\includegraphics[scale=0.5]{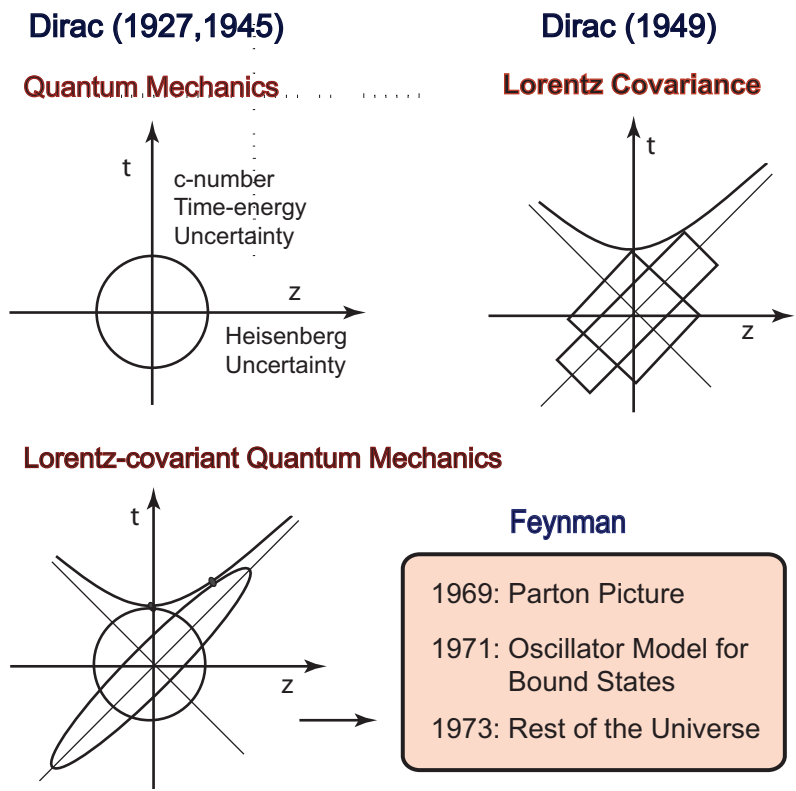}}
\vspace{1mm}
\caption{Dirac's quantum mechanics (1927,1945) and Dirac's relativity (1949).
If they are combined, it leads to a Lorentz-covariant bound-state picture
which produces the quark and parton models are two different limiting
cases of one formula, just as in the case of Einstein's energy-momentum
formula.}\label{fig.5}
\end{figure}

\par

For bound-state problems, Paul A. M. Dirac wrote three important papers
on this subject (Dirac 1927,1945, 1949).  His 1927 paper tells us there
is a time-energy uncertainty relation.  In 1945, he attempted to use the
harmonic oscillator to formulate quantum mechanics applicable
to Einstein's world.  If we combine or Hegelize Dirac's 1927 and 1945
papers, we end up with the circle given in Fig.~5.

\par
In 1949, Dirac showed
the Lorentz boost can be described as a squeeze transformation as shown
in Fig.~5.  If we Hegelize the circle and the squeezed rectangle,
we arrive at the ellipse (Kim and Noz 2011) which can explain what
happens in the real world including the quark model (Gell-Mann 1964)
and the parton model (Feynman 1969).  This Hegelian procedure corresponds
to Step 1 in Fig.~4.
\par
The final step in constructing Lorentz-covariant quantum mechanics
is to show that the scattering and bound states share the same set
of fundamental principles (Han et al. 1981).   This Hegelians procedure
is illustrated in as Step 2 in Fig.~4.

\par
\begin{figure}
\centerline{\includegraphics[scale=2.5]{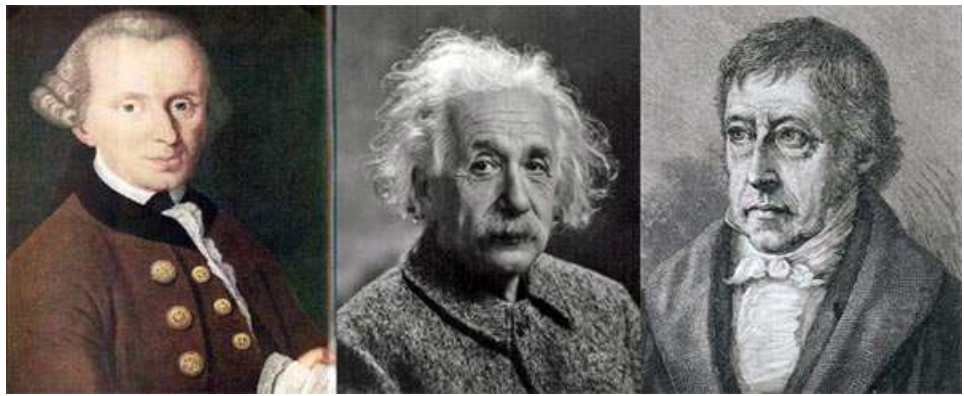}}
\vspace{1mm}
\caption{Kant, Einstein, and Hegel.  It is appropriate to place
Einstein between Kant and Hegel.}\label{fig.6}
\end{figure}

\par

\section{Kant, Hegel, and Einstein}
Kant and Hegel are two of the most fundamental thinkers affecting
our present-day lifestyles.  However, their philosophies were based
largely on social events and applicable to formulation of social
sciences.  It is gratifying to note that Einstein gives us a more
concrete picture of their approaches to the problems.  By building
the bridge between Kant and Hegel as illustrated in Fig.~6, Einstein
not only gives us the precise description of how physical theories
were developed in the past but also tells us how to approach current
problems in physics.

\end{document}